\documentclass[aps,pra,reprint,superscriptaddress]{revtex4-1}

\usepackage{graphicx}
\usepackage{color}
\usepackage{amsmath,amssymb}
\usepackage{bm}
\usepackage{upgreek}
\usepackage{xspace}
\usepackage[colorlinks,urlcolor=blue,citecolor=blue,linkcolor=blue]{hyperref}
\graphicspath{{figures/}}

\newcommand{\ket}[1]{\vert #1 \rangle \xspace}

\begin{document}

\title{Interplay between geometric and dynamic phases in a single spin system}

\author{A.~A.~Wood}
\email{alexander.wood@unimelb.edu.au}
\affiliation{School of Physics, University of Melbourne, Victoria 3010, Australia}
\author{K.~Streltsov}
\affiliation{Institute for Theoretical Physics, Ulm University, Germany}
\author{R.~M.~Goldblatt}
\affiliation{School of Physics, University of Melbourne, Victoria 3010, Australia}
\author{M. B. Plenio}
\affiliation{Institute for Theoretical Physics, Ulm University, Germany}
\author{L. C. L. Hollenberg}
\affiliation{School of Physics, University of Melbourne, Victoria 3010, Australia}
\affiliation{Centre for Quantum Computation and Communication Technology, University of Melbourne, Victoria 3010, Australia}
\author{R. E. Scholten}
\affiliation{School of Physics, University of Melbourne, Victoria 3010, Australia}
\author{A. M. Martin}

\affiliation{School of Physics, University of Melbourne, Victoria 3010, Australia}

\date{\today}
\begin{abstract}
We use a combination of microwave fields and free precession to drive the spin of a nitrogen-vacancy (NV) center in diamond on different trajectories on the Bloch sphere, and investigate the physical significance of the frame-dependent decomposition of the total phase into geometric and dynamic parts. The experiments are performed on a two-level subspace of the spin-1 ground state of the NV, where the Aharonov-Anandan geometric phase manifests itself as a global phase, and we use the third level of the NV ground state triplet to detect it. We show that while the geometric Aharonov-Anandan phase retains its connection to the solid angle swept out by the evolving spin, it is generally accompanied by a dynamic phase that suppresses the geometric dependence of the system dynamics. These results offer insights into the physical significance of frame-dependent geometric phases.     
\end{abstract}
\maketitle

When a quantum system evolves in a manner that can be represented geometrically -- such as a spin-1/2 qubit on the Bloch sphere -- a phase that depends on the geometry of the evolution arises in addition to the `dynamic' phase accumulated as a result of time evolution. Such `geometric' phases form a unified framework for interpreting the evolution of a quantum system~\cite{anandan_geometric_1992}. First identified by Pancharatnam~\cite{pancharatnam_generalized_1956} and later studied by Berry~\cite{berry_quantal_1984} for adiabatically-varying periodic Hamiltonians, it was realised by Aharonov and Anandan~\cite{aharonov_phase_1987} that cyclic \emph{state} evolution itself also results in a geometric phase. Aharonov and Anandan's phase is the gauge-invariant generalisation of Berry's phase in the adiabatic limit, while further work generalized the idea of geometric phase to non-adiabatic, non-cyclic evolution~\cite{samuel_general_1988}. A key aspect of Aharonov and Anandan's work was the explicit identification of the dynamic and geometric components of the overall phase acquired for a cyclically-evolving quantum state. This  decomposition expresses the unique nature of the geometric phase, while at the same time underlining the difficulty in detecting it independently of the dynamic phase.  

 The advent of solid-state, single-quantum spin systems has triggered renewed interest in geometric phase, with notable measurements using superconducting qubits ~\cite{leek_observation_2007, abdumalikov_jr_experimental_2013} and spin qubits in diamond~\cite{zu_experimental_2014, arroyo-camejo_room_2014}. Nitrogen-vacancy (NV) centers in diamond~\cite{doherty_nitrogen-vacancy_2013, schirhagl_nitrogen-vacancy_2014, wu_diamond_2016} offer a particularly attractive system to study and utilize geometric phases, featuring a ground-state spin triplet with millisecond coherence times~\cite{balasubramanian_ultralong_2009} at room temperature, quantum states amenable to microwave geometric gates~\cite{zhang_experimental_2016}, optical fields~\cite{yale_optical_2016, zhou_holonomic_2017} and even physical rotation~\cite{maclaurin_measurable_2012, ledbetter_gyroscopes_2012}. A key factor in this renewed interest is the potential to exploit features of the geometric phase that distinguish it from the dynamic phase, such as resilience to certain kinds of noise, which has application in quantum computation~\cite{zanardi_holonomic_1999}. Although geometric phases are a ubiquitous feature of quantum systems, whether or not geometry in a given setting confers unique properties to the system, or simply offers an alternative formulation of the dynamics, is often not clear.   

In this work, we experimentally investigate the Aharonov-Anandan non-adiabatic geometric phase using distinct two-level subspaces of a single NV ground state manifold. We can individually address subspaces due to the significant frequency difference between the $m_S = 0\rightarrow +1$ and $m_S = 0\rightarrow -1$ transitions in the NV ground state triplet when an external magnetic field is applied (Fig. \ref{fig:fig1}(a, b)). The Aharonov-Anandan phase is defined for cyclic evolutions in projective Hilbert space, and therefore manifests itself as a global phase in the two-level subspace. When considering the additional third level of the ground state triplet this global phase can nevertheless be detected as described in the following Section. We drive the NV spin along cyclic trajectories by applying sequences of detuned microwave pulses and free precession dynamics. We show that the interplay between dynamic and geometric phase in this system results in the experimentally-measured phase shift being invariant to geometric manipulations. These results reveal the subtle nature of the Aharonov-Anandan phase and the significance of dynamic phases in a quantum spin system.    

This paper is structured as follows. In Section \ref{sec:geo}, we review the Aharonov-Anandan formulation of geometric and dynamic phases, and consider specifically the theory pertinent for examining these in the context of the NV center in diamond. Section \ref{sec:exp} presents an overview of our experimental scheme and methods, and explores the significance of a global phase accumulated by a standard spin-echo interferometry sequence. In Section \ref{sec:exp2} we demonstrate and characterize microwave pulses that drive the NV spin on closed trajectories in Hilbert space. In Section \ref{sec:comp} we combine circuit-driving microwave pulses with free evolution in the rotating frame to understand the significance of frame-dependent geometric phases. Section \ref{sec:disc} discusses these results and offers an outlook on our results as well as further avenues of research. 

\section{Geometric phases in quantum spin systems \label{sec:geo}}

Confirmation and demonstration of geometric phase in numerous systems and settings~\cite{wilczek_geometric_1989} were the focus of experimental work immediately following Berry's initial realization~\cite{berry_quantal_1984}. As Berry's phase depends only on the geometry of a given evolution, this renders it insensitive to certain forms of noise, sparking interest as a potentially fault-tolerant quantum gate operation~\cite{jones_geometric_2000, de_chiara_berry_2003}. However, as Berry's phase demands adiabatic evolution, which translates to slow quantum gate operations, environmental decoherence becomes a serious problem given the short coherence time of many solid-state quantum systems~\cite{carollo_geometric_2003}. Attention then turned to non-adiabatic geometric phases, where the requirement for slow evolution is relaxed, but the presence of dynamic phases that accompany the spin manipulations used to drive geometric gates reintroduces a conduit for noise to enter the system. Several schemes have been investigated to remove the dynamic phase, which typically cannot be eliminated in a time-reversal measurement such as spin-echo without also removing the geometric phase~\cite{wang_NMR_2001, zhu_universal_2003,ota_geometric_2009}. 

The often unavoidable interplay between dynamic and geometric phase underpins the difficulty in realizing the purported advantages of geometric phase manipulations on spin qubits. The distinction between dynamic and geometric phase provoked work that suggested the geometric phase can be formally removed and absorbed into the dynamic phase, yet still retain its `geometric character'~\cite{giavarini_removing_1989}. Other work concerned the invariance of geometric phases to unitary transformations~\cite{kobe_invariance_1990, kendrick_invariance_1992}. 

A simple intuition for the geometric nature of the geometric phase in two level systems can be gained by expressing its relationship to the solid angle $\Theta$ enclosed by the traversed path of the state in the associated space, namely

\begin{equation}\label{eq:rob}
\Phi_G = m\Theta,
\end{equation} 

with $m$ the spin quantum number. This space can always be represented by a sphere: for the adiabatic Berry phase this is the space of parameters describing the Hamiltonian, the Poincare sphere that parametrizes light polarisations is the relevant space for Pancharatnam's geometric phase. For a beam of light directed through different spatial directions, the relevant parameter space is simply the unit sphere~\cite{chiao_observation_1988, ben-aryeh_berry_2004}. The Bloch sphere, being the standard representation of a two-level quantum system, is also a common parameter space whereby a geometric phase is defined. The Bloch sphere is an example of a \emph{projective} Hilbert space: quantum states with different overall (global) phases are mapped to the same point on the Bloch sphere.

\subsection{The Aharonov-Anandan phase}

The notion of a quantum system evolving in projective Hilbert space is at the core of Aharanov and Anandan's 1987 work~\cite{aharonov_phase_1987}. For a Hamiltonian $H$, a state vector $\ket{\psi(t)} = \exp(-iHt)\ket{\psi(0)}$ ($\hbar = 1$) undergoes evolution for a time $T$ such that it returns to its initial configuration at $t = T$ with a global phase difference, $\ket{\psi(T)} = \exp(i\phi_T)\ket{\psi(0)}$. Aharanov and Anandan showed that when mapped to a projective Hilbert space, \emph{i.e.} $\ket{\tilde{\psi}} = e^{-if(t)}\ket{\psi}$, $f(T)-f(0) = \phi$ and thus $\ket{\tilde{\psi}(T)} = \ket{\tilde{\psi}(0)}$, the total phase $\phi$ can be expressed as
\begin{equation}\label{eq:aap}
\phi(T) = -\int_0^T\langle\psi\vert H \vert\psi\rangle\,dt+\int_0^T i\langle\tilde{\psi}\vert \frac{d}{dt} \vert\tilde{\psi}\rangle\,dt,
\end{equation}
where the first term is identified as the dynamic phase $\phi_\text{dyn}$, and the second term as the geometric or Aharanov-Anandan (AA) phase, $\phi_\text{AA}$. 

Since the AA phase concerns cyclic evolutions of a quantum state, the initial and final states on the Bloch sphere are the same, differing only by an overall phase. This poses a challenge to experimentally detecting the AA phase. The first experimental demonstration of Aharonov and Anandan's geometric phase was by Suter \emph{et al.} in 1987~\cite{suter_study_1988}. A spin-1 system of coupled protons served as a spin-1 system that could be decomposed into two independently addressable spin-1/2 systems due to dipolar coupling rendering the frequency of each transition different. The spin in one two-level subspace which was driven about closed circuits on the Bloch sphere, accumulating a global geometric phase, and the third state served as a reference interferometer to detect the phase picked up by the first system. An in-depth theoretical analysis of this experiment was the topic of Ref. \cite{skrynnikov_geometric_1994}. We will make extensive use of the idea that in the NV system, geometric phases in a spin-1/2 subspace of a fundamentally spin-1 quantum system can be detected using reference interferometry.

\subsection{Application to the NV spin: global phases}

\begin{figure}
	\centering
		\includegraphics[width = \columnwidth]{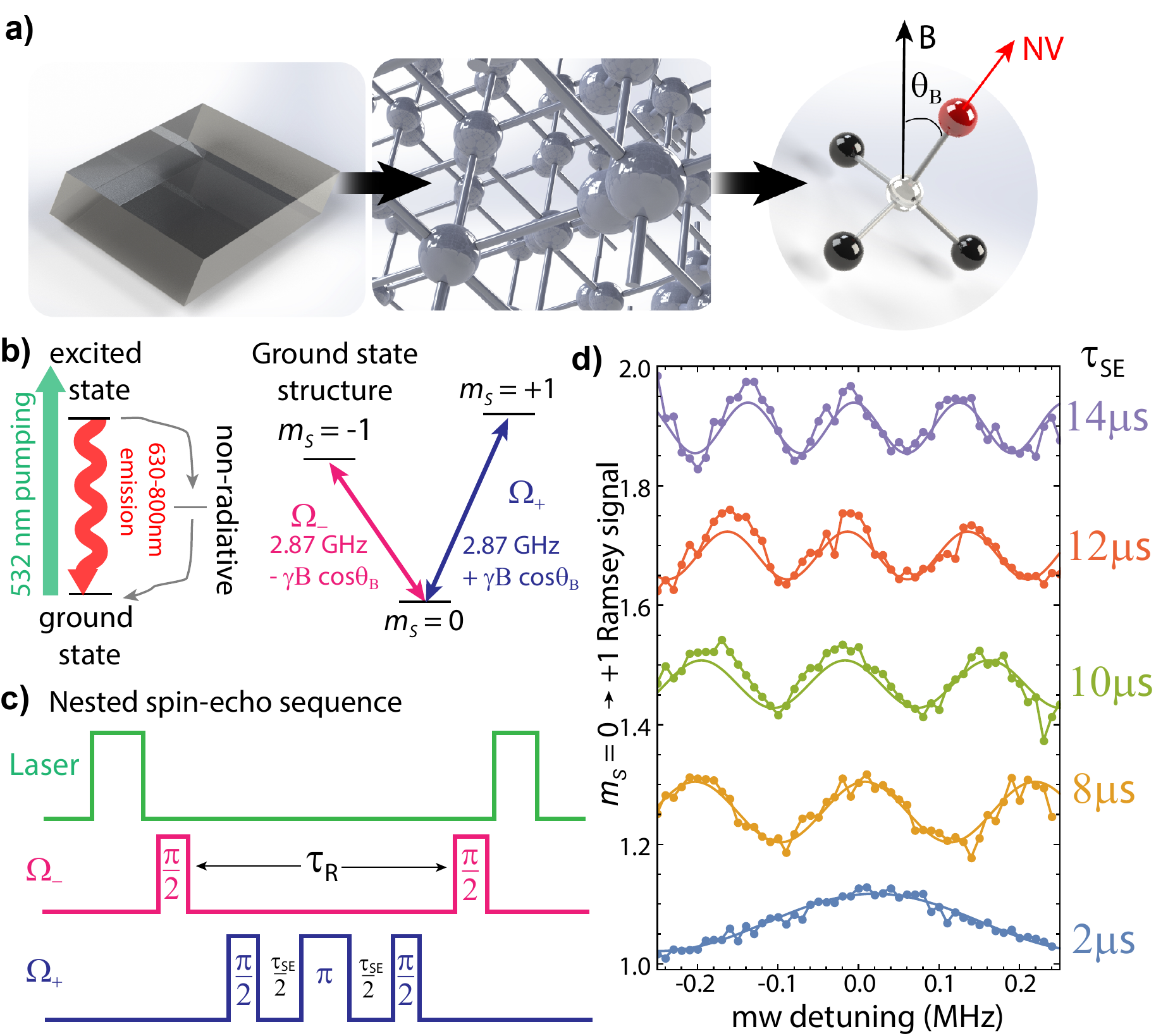}
	\caption{(a) the NV center in diamond, formed from a subsitutional nitrogen adjacent to a lattice vacancy in the diamond matrix. (b) Optical transitions and ground state structure of the NV, depicting Zeeman-separated $m_S$ energy levels and transition frequencies. (c) A variable-detuning spin-echo sequence nested inside a resonant Ramsey sequence, showing microwaves applied to each two-level subspace and laser preparation and readout pulses. The time between spin-echo pulses $\tau_\text{SE}$ is varied and time delays either side of the spin echo sequence are added to ensure the total time between the Ramsey pulses $\tau_R$ remains fixed. The resonant microwave Rabi frequency for each transition is about $500\,$kHz. (d) Ramsey fringes for different spin-echo times $\tau_\text{SE}$ from the $\{0,-1\}$ interferometer while only the detuning $\Delta$ of the $m_S = 0\rightarrow m_S = -1$ microwaves are changed. The frequency of the Ramsey fringes is consistent with $\Delta \tau_\text{SE}/4$, albeit slightly offset due to the low Rabi frequency effectively amplitude modulating the interference fringes. This shows that spin-echo results in a global phase in the $\{0,+1\}$ subspace that can be read out using interferometry on the $m_S = 0\rightarrow m_S = -1$ transition. 
 }
	\label{fig:fig1}
\end{figure}

The room-temperature level structure of the NV center triplet ground state, depicted in Fig. \ref{fig:fig1}(a,b) derives from strong spin-spin coupling between two electrons~\cite{loubser_electron_1978} and a Zeeman shift in the presence of a magnetic field component $B$ parallel to the N-V axis:
\begin{equation}
H = D_\text{zfs} S_z^2 + \gamma B S_z
\end{equation}
with $D_\text{zfs}/2\pi = 2.870\,\text{GHz}$ NV zero-field splitting, $S_z$ the spin-1 $z$ Pauli matrix and $\gamma/2\pi = 2.8\,\text{MHz\,G}^{-1}$.  For low magnetic field strengths ($B \ll 1000\,$G), the zero-field splitting is the dominant interaction. We consider now the application of two microwave fields $\boldsymbol{\omega}_\pm(t)$, tuned near resonance of the $m_S = \pm1$ transitions, with respective resonant Rabi frequencies $\Omega_\pm$, angular frequencies $\omega_\pm$ and phases $\phi_\pm$, the subscripts $\pm$ denoting the respective $m_S = \pm1$ basis, see Fig. \ref{fig:fig1}(b). The Hamiltonian for the driven NV center in the composite frame rotating at each microwave frequency is given by \cite{mamin_multipulse_2014}
\begin{equation}
H_{\text{int}} = \left(
\begin{matrix}
-\Delta_+&\frac{\Omega_+ e^{-i\phi_+}}{2\sqrt{2}}&0\\ 
\frac{\Omega_+ e^{i\phi_+}}{2\sqrt{2}}&0&\frac{\Omega_-e^{-i\phi_-}}{2\sqrt{2}}\\ 
0&\frac{\Omega_- e^{i\phi_-}}{2\sqrt{2}}&-\Delta_-
\end{matrix}
\right),\label{eq:ham}
\end{equation}
with $\Delta_+ = \omega_+ - D_\text{zfs} + \gamma B $, $\Delta_- =\omega_- - D_\text{zfs} - \gamma B $. We assume also that $\Delta_\pm \gg \Omega_\pm$, which facilitates dropping $\Omega_\mp$ terms that appear in the $m_S =\pm1$ subspaces. We introduce effective spin-1/2 Hamiltonians that conveniently describe the coupling between the NV spin and a microwave field for the $\{m_S = 0, m_S = +1\}\equiv\{0,+1\}$ and$\{m_S = 0, m_S = -1\}\equiv\{0,-1\}$ subspaces,
\begin{equation}
H_+ = \left(
\begin{matrix}
-\Delta_+ &\frac{\Omega_+ e^{-i\phi_+}}{2}\\
\frac{\Omega_+e^{i\phi_+}}{2} & 0
\end{matrix}
\right),
H_- = \left(
\begin{matrix}
0 &\frac{\Omega_-e^{-i\phi_-}}{2}\\
\frac{\Omega_-e^{i\phi_-}}{2} & -\Delta_-
\end{matrix}
\right).\label{eq:h2}
\end{equation}

A demonstrative example of a global phase measurement using the three levels of the NV spin is that of a spin-echo sequence addressing one particular subspace, for instance, the 2-level ${0,+1}$ subspace:
\begin{equation}
\ket{\psi(\tau)} = U_{+,\pi/2}U_\text{free}U_{+,\pi}U_\text{free}U_{+,\pi/2}\ket{\psi_0},\label{eq:se}
\end{equation}
with $U_{+,\pi/2}$ and $U_{+,\pi}$ the $\pi/2$-pulse and $\pi$-pulse operators (with $\Omega_- = 0$) and the free-evolution operator $U_\text{free}(\tau) = \exp(-i H t)|_{\Omega_+\rightarrow 0}$. For the spin-echo sequence Eq. \ref{eq:se} with $\Omega_- = 0$ and the NV initialized into the $m_S = +1$ state, $\ket{\psi_0} = \ket{+1}$, we find 
\begin{equation}
\ket{\psi(\tau)} = -e^{i \Delta_+ \tau/2}\ket{\psi_0}.
\end{equation}
The spin-echo sequence does not \emph{cancel} the DC detuning $\Delta_+$, it maps it to a global phase on the $\{0,+1\}$ subspace. We can modify the initial state so that the global phase manifests as a relative phase between states by first applying a resonant ${0,-1}$ $\pi/2$-pulse to an initially pure $m_S = 0$ state, and a final ${0,-1}$ $\pi/2$-pulse on the $m_S = -1$ basis to close an enveloping Ramsey interferometry sequence,
\begin{equation}
\ket{\psi(\tau)} = \underbrace{U_{-,\pi/2}\overbrace{U_{+,\pi/2}U_\text{free}U_{\pi}U_\text{free}U_{+,\pi/2}}^{\text{spin-echo}}U_{-,\pi/2}}_{\text{Ramsey}}\ket{\psi_0}.\label{eq:se2}
\end{equation}
We find $\ket{\psi(\tau)} = -(1+e^{\frac{i\Delta_+ \tau}{2}})/2\ket{0}+i(1-e^{\frac{i\Delta_+ \tau}{2}})/2\ket{+1}$, the NV $m_S = 0$ bright state population then varies as $\cos^2\left(\Delta_+\tau/4\right)$. We can compare this result to a standard two-level system representing only one NV subspace with Hamiltonian $H_\text{2LS} = -\Delta/2\sigma_z+\Omega/2\sigma_x$ ($\phi = 0$), $\sigma_{x,y,z}$ the standard spin-1/2 Pauli matrices. For an initial spin state in the spin-1/2 basis of $H_\text{2LS}$, $\ket{\psi_{(2\text{LS})}(0)} = \ket{\uparrow}$, application of a spin-echo sequence yields
\begin{equation}
\ket{\psi_{(2\text{LS})}(\tau)} = -\ket{\uparrow}
\end{equation}
for all $\Delta$, the global phase is now apparently independent of $\tau$, in contrast to the spin-1 result. However, using the two-level Hamiltonians defined in Eq. \ref{eq:h2} yields the correct global phase for a spin-echo sequence, we therefore will also use Eq. \ref{eq:h2} for calculating the dynamic and Aharonov-Anandan phases within the pseudospin-1/2 subspaces. A consequence of this choice of Hamiltonian are geometric and dynamic phases that no longer obey the conventional spin-1/2 relations~\cite{wang_NMR_2001} defining the contribution of each phase to the overall phase. As we shall see, this has no bearing on the measured outcome of each experiment, but is an important topic of discussion when interpreting the results of a given pulse sequence, reserved for Section \ref{sec:disc}. 

Spin-echo represents an archetypal case of cyclic evolution of a state vector leading to a global phase. However, the relative contributions of geometric and dynamic phase are not obvious -- an intuitive picture would lead one to conclude that the path the state vector traces out during the spin-echo sequence subtends a solid angle, and that in turn defines the geometric phase. If this were the case, one would also expect a rather more complicated global phase than simply $\phi = \Delta_+ \tau/2$ given the application of a a $\pi$-pulse halfway into the free evolution time, that subtends a $\tau$-dependent spherical cap. As we shall see, the total global phase is not purely geometric: the decomposition of geometric and dynamic phase is more involved when evolution occurs over several different Hamiltonians. We now proceed to describe the experimental realization of reference interferometry with NV centers in diamond and measurements of global phases.
  
\section{Experiment}
\label{sec:exp}

Our experimental setup is designed to rotate diamonds containing single NV centers~\cite{wood_quantum_2018, wood_observation_2020} and consists of a $^{12}$C-enriched CVD diamond mounted on its (100) face. The motor that ordinarily rotates the diamond is held constant at a static rotation angle and not altered during any experiments. A purpose-built confocal microscope is used to optically address and read single NV centers with typical coherence time $T_2^\ast = 50\,\upmu$s, $T_2 =1\,$ms. A 15\,G magnetic bias field oriented along the surface normal of the diamond ($\theta = 54.7^\circ$ to the NV axis) is used to break the degeneracy of the $m_S = \pm1$ states \footnote{The orientation of the bias magnetic field in this case is a consequence of technical limitations inherent to our experiment, and has no bearing on the results of this work.} and a $20\,\upmu$m diameter copper wire $100\,\upmu$m above the diamond surface is used to apply linearly-polarized microwave fields tuned to the $|0\rangle\rightarrow|\pm1\rangle$ transitions. We operate with microwave fields tuned to the $|m_S=\pm1, m_I = 0\rangle$ hyperfine transition (nuclear spin quantum number is conserved) and ensure low microwave Rabi frequencies ($\Omega\leq500\,$kHz) so that off-resonant dressing of the proximal $m_I = \pm1$ hyperfine states is minimized~\cite{arai_geometric_2018}. Microwaves are generated from three independently tunable sources, gated using fast switches and combined before being amplified and applied to the NV center. The microwave pulses operating on each two-level subspace are never applied simultaneously.     

\subsection{Nested spin-echo}
We first confirm that a static microwave detuning yields a global phase on the final state of a spin-echo sequence targeting a pseudospin-1/2 subspace of the NV, as discussed in the previous Section. The pulse sequence we use is depicted in Fig. \ref{fig:fig1}(c), and consists of a resonant Ramsey sequence applied to the $\{0,-1\}$ subspace while a spin-echo sequence with pulse detuning $\Delta$ is applied to the $\{0, +1\}$ subspace~\cite{rong_enhanced_2011}. The complete sequence begins and ends with a 3$\upmu$s laser pulse that prepares and reads the NV spin. Considered within each subspace individually, the signal resulting from the spin-echo sequence and the Ramsey sequence are unaffected by varying the spin-echo pulse detuning. However, from Fig~\ref{fig:fig1}(b), we see that spin-echo actually yields a measurable phase that is read out using the Ramsey sequence. We will return to a discussion of spin-echo in the context of the results of Section \ref{sec:comp}, which explore the significance of geometric phases under cyclic evolutions driven by multiple Hamiltonians. We will now move to studying microwave pulses that drive cyclic evolutions on the Bloch sphere that result in geometric phases.

\section{Measurement of geometric and dynamic phases}\label{sec:exp2}
\begin{figure}
	\centering
		\includegraphics[width = \columnwidth]{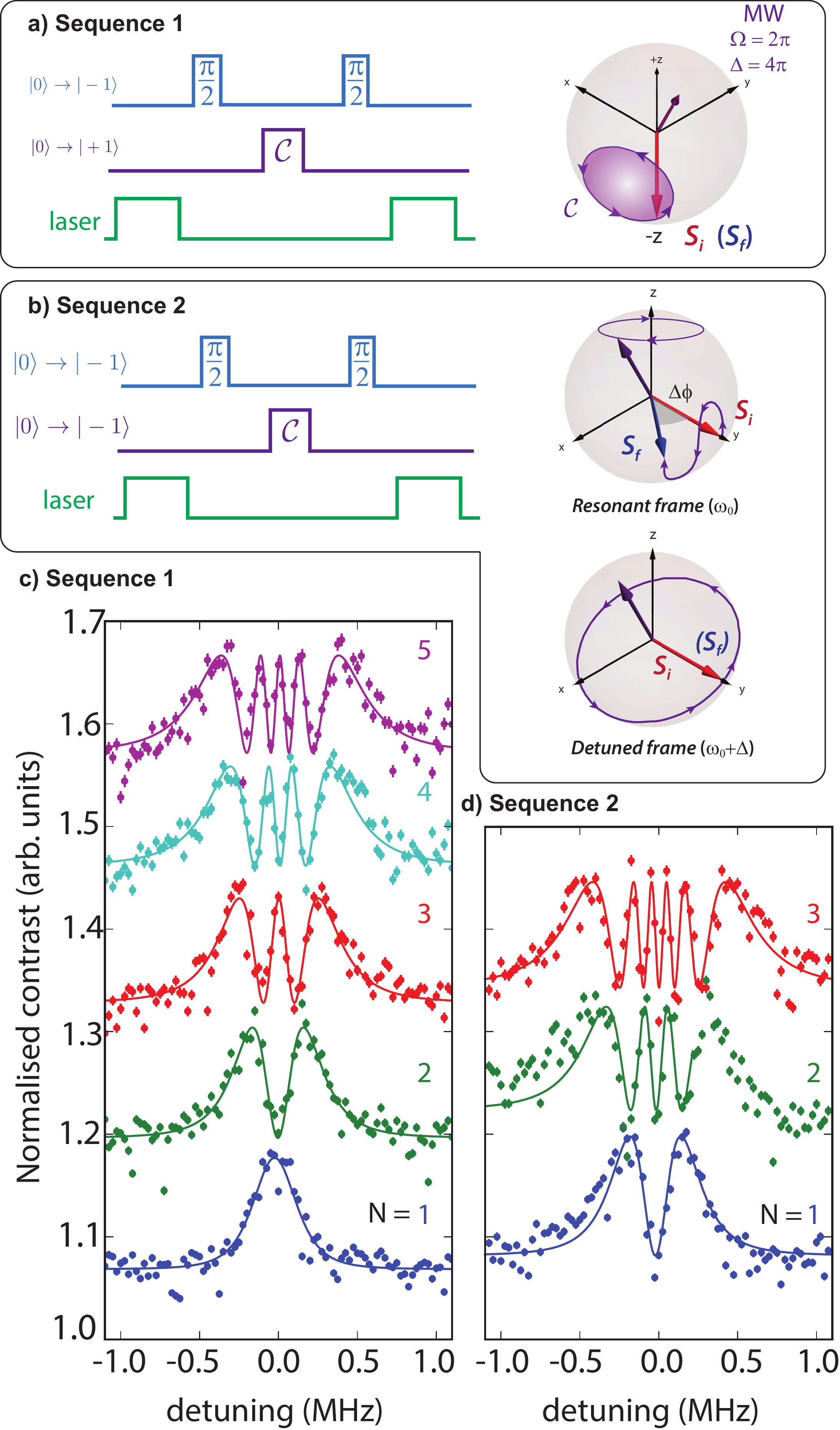}
	\caption{Pulse sequences applied to the NV spin to drive cyclic trajectories and measured spin populations for multiple circuits. (a) Sequence 1: a resonant Ramsey sequence applied to the $\{0,-1\}$ subspace reads out the phase accumulated by the $\mathcal{C}$-pulse applied to the $\{0,+1\}$ system. The path traversed by the spin in the Bloch sphere consists of a closed circle subtending a cone with solid angle $\Theta$. b) Sequence 2, both the resonant interferometry sequence and $\mathcal{C}$-pulse act on the $\{0,-1\}$ system. The path the spin takes in the resonant frame is non-cyclic, with initial and final azimuthal spins separated by a relative phase $\Delta\phi$. Transformation to the detuned frame of reference yields a cyclic evolution, though the geometric and dynamic phases both depend on the absolute phase of the $\mathcal{C}$-pulse, which is uncontrolled. (c,d) Experimental results for varying the $\mathcal{C}$-pulse detuning and cumulative rotation number $N$ for Sequences 1 and 2, respectively. Error bars derive from photon counting statistics and lines are fits of the form $\cos^2(N\phi)$.} 
	\label{fig:fig2}
\end{figure}

In this Section, we drive the NV spin using detuned microwave (mw) pulses timed to execute a cyclic evolution of the spin, which in turn induces a geometric phase. For a state initially parallel to to the $+z$-axis, application of the mw field for one Rabi period $t_{2\pi}(\Delta) =2\pi/\sqrt{\Omega^2+\Delta^2}$ returns the quantum state to its initial state, executing a cone trajectory on the Bloch sphere, Fig. \ref{fig:fig2}(a). We therefore call this a $\mathcal{C}$-pulse, and henceforth drop the $\pm$ subscripts for each microwave field, since we only ever vary the detuning in a single specified subspace. We define the solid angle swept by the spin vector during the evolution as
\begin{equation}
\Theta = 2\pi\left(1-\frac{\boldsymbol{\Omega}\cdot\ \boldsymbol{S}}{|\boldsymbol{S}||\boldsymbol{\Omega}|}\right)\label{eq:solidangle}
\end{equation}
with $\boldsymbol{S} = \langle\sigma_i\rangle$, $i\in\{x,y,z\}$ the Bloch vector of the spin state given in Eq. \ref{eq:sv} and $\boldsymbol{\Omega}$ satisfies $d\boldsymbol{S}/dt =\boldsymbol{S}\times\boldsymbol{\Omega}$ for a given $\boldsymbol{S}$; in the $\{0,+1\}$ subspace $\boldsymbol{\Omega} = (-\Omega_0, 0, \Delta)$\footnote{This definition is consistent with the original formulation by Bloch, where the spin vector precesses clockwise around the effective magnetic field $\boldsymbol{\Omega}$, see F. Bloch, \href{https://doi-org.ezp.lib.unimelb.edu.au/10.1103/PhysRev.70.460}{Phys. Rev. {\bf 70} 460 (1946)} and J. D. Roberts, \href{ https://doi.org/10.1002/cmr.1820030104}{Concept. Magn. Resonan. {\bf 3} 27-45 (1991)}.}. As a simple example, consider a $\mathcal{C}$-pulse applied to the initially spin-down state, $\ket{0}$ in the $\{0,+1\}$ subspace. Then $\Theta = 2\pi(1 + \Delta/\sqrt{\Delta^2+\Omega_0^2})$, for $\phi = 0$, and the AA phase for this path is then
\begin{equation}
\phi_\text{AA} = \pi\left(1+\frac{\Delta}{\sqrt{\Delta^2+\Omega^2}}\right).
\label{eq:aasa}
\end{equation}
We note for this path, $\ket{\psi(t_{2\pi})} =e^{i\phi_T}\ket{0}$, with $\phi_T = \pi(1+\Delta/\sqrt{\Delta^2+\Omega^2}) = \phi_\text{AA}$, which implies zero dynamic phase.

Multiple cyclic evolutions are obtained by applying the driving field for an integer multiple $N$ of $t_{2\pi}$, in which case the AA phase is given by $N\phi_\text{AA}$, and the NV $m_S = 0$ population varies as $\cos^2(N\phi_{AA})$. We study several pulse sequences applied to drive the NV spin along the `cone' trajectory which differ by the initial configuration of spin states and which two-level subsystem is driven by the $\mathcal{C}$-pulse, as shown in Fig. \ref{fig:fig2}. A $\mathcal{C}$-pulse is applied to either the $\{0,+1\}$ (Sequence 1) or $\{0,-1\}$ subspace (Sequence 2) after an initially pure $\ket{0}$ state is put into a superposition with the $|-1\rangle$ state by a resonant $\pi/2$-pulse. We vary the cone solid angle by changing the detuning of the $\mathcal{C}$-pulse, and the resulting phase accumulation is read out by a final resonant $\pi/2$ pulse, closing the Ramsey interferometry sequence operating on the $\{0,-1\}$ subspace. The results for experiments using Sequences 1 and 2 is shown in Fig. \ref{fig:fig2}(c,d), for increasing values of $N$. We compared our experimental results to numerical simulations of the three-level NV spin system subjected to resonant and detuned driving fields, and found satisfactory agreement.

Sequence 1 corresponds to the archetypal cone trajectory considered by Ref. \cite{suter_study_1988} where the AA phase manifests as a global phase on the $(|0\rangle\rightarrow|+1\rangle)$ system. Sequence 2 is similar, except that both the interferometry pulse sequence ($\pi/2$-pulses) and $\mathcal{C}$-pulses act on the same two-level subsystem, and measurement of a relative phase $\Delta\phi$ by the interferometry sequence immediately suggests the path the $\mathcal{C}$-pulse executes is \emph{not} cyclic in its subspace, since this relative phase maps to an azimuthal separation between initial and final states on the Bloch sphere (Fig. \ref{fig:fig2}b).

The phase of the driving field also plays an important role for Sequence 2. The initial spin vector acted on by the $\mathcal{C}$-pulse is an azimuthal vector on the Bloch sphere with a well-defined angle with respect to the $x-y$ axes, namely, the phase of the microwave driving field that executed the first $\pi/2$-pulse. In our experiments, different microwave sources are used to drive the resonant $\pi/2$-pulses and off-resonant $\mathcal{C}$-pulses, meaning there is no phase relationship between the two driving fields, even when driving spins within the same subspace. Therefore, the angle between the microwave driving field used to drive the $\mathcal{C}$-pulse and the azimuthal spin is effectively random on a shot-to-shot basis: the corresponding solid angle, and thus $\phi_\text{AA}$, is also random. Since Sequence 1 applies a $\mathcal{C}$-pulse to an initially longitudinal spin state, the relative phase dependence is suppressed, the detuning of the microwave field being the only parameter determining the solid angle.

\begin{figure}
	\centering
		\includegraphics[width = \columnwidth]{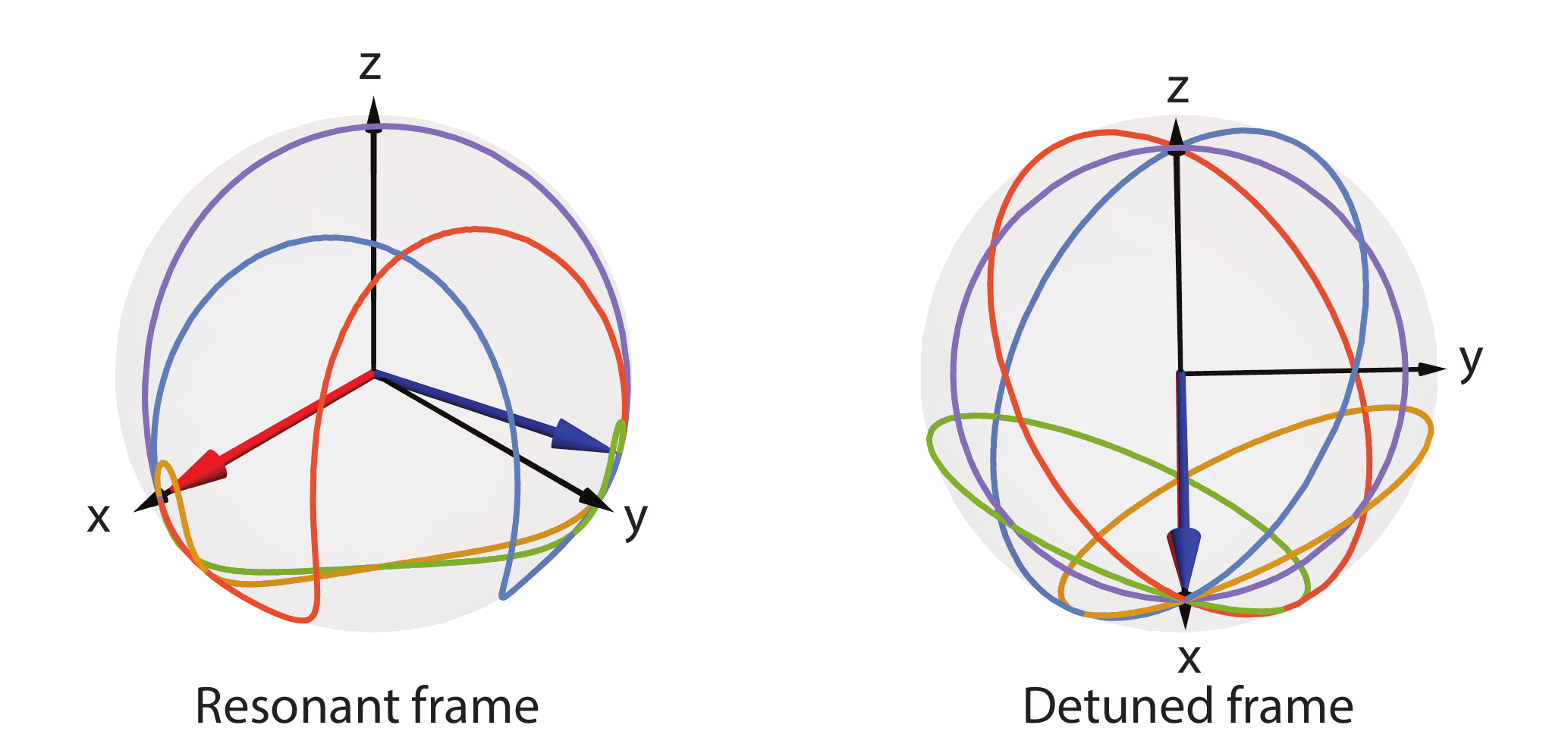}
	\caption{Calculated paths taken by the Bloch vector of the NV spin under the application of Sequence 2, observed in the \emph{resonant frame} (left), rotating at the two-level splitting (equivalently, the resonant microwave frequency) and (right) in the \emph{detuned frame}, rotating at $\omega\pm\Delta$, where $\Delta$ is the detuning of the $\mathcal{C}$-pulse. Different colors represent different phases of the microwave field sampled when the $\mathcal{C}$-pulse is applied. In the resonant frame, where the NV spin Bloch vector is stationary, the $\mathcal{C}$-pulse drives non-cyclic trajectories, so that the initial and final spins are separated azimuthally by $\Delta\phi$. In the detuned frame, the same trajectories are closed, subtending solid angles proportional to the Aharonov-Anandan phase.}
	\label{fig:fig3}
\end{figure}

While the trajectory in Sequence 2 executed by the state vector during the $\mathcal{C}$-pulse is not cyclic in the frame rotating at the resonant frequency (the frame in which the interferometry sequence is performed), it is cyclic in the frame rotating at the frequency of the $\mathcal{C}$-pulse. In this frame, the gauge transformation $\ket{\tilde{\psi}} = e^{-if(t)}\ket{\psi}$ with $f(t) = t/2(\Delta + \sqrt{\Delta^2+\Omega^2}$ yields $\ket{\tilde{\psi}(t_{2\pi})} = \exp(-i\phi_T)\ket{\tilde{\psi}(0)}$, with $\phi_T = \phi_\text{dyn}+\phi_\text{AA}$. The spin vector is rotated a full cycle about the microwave field $\boldsymbol{\Omega}$, even though the relative azimuthal angle between the spin vector and the microwave field is still an effectively random quantity. Figure \ref{fig:fig3} shows exemplar paths taken by the spin vector in the resonant frame and detuned frames, with several non-cyclic trajectories in the resonant frame all returning to the same point in the detuned frame and subtending solid angles proportional to the Aharonov-Anandan phase.

In the detuned frame, application of Eq. \ref{eq:aap} yields
\begin{subequations}
\begin{eqnarray}
\phi_\text{dyn} =& \pi\left(\Delta -\Omega\,\sin(\phi_0)\right)/\sqrt{\Delta^2+\Omega^2}\\
\phi_\text{AA} =& \pi\left(1+\Omega\,\sin(\phi_0)/\sqrt{\Delta^2+\Omega^2}\right)
\label{eq:faa}
\end{eqnarray}
\end{subequations}
and the total phase is also $\phi_T  = \pi\left(1+\Delta/\sqrt{\Delta^2+\Omega^2}\right) = \Delta\phi$, the phase difference measured by the Ramsey interferometry sequence performed on the system. The microwave-phase-dependent terms of the dynamic and geometric terms cancel, and since the relative phase between the states of the two-level system is equal and opposite, the accumulated phase is twice that of Sequence 1. 

As a demonstrative example of Aharonov and Anandan's formulation of non-adiabatic geometric phases, the results of Sequences 1 and 2 highlight some important features of the AA phase in a spin-1 system measured in a rotating reference frame. We first examine the non-adiabaticity: in the resonant frame, where the spin vector is stationary, the microwave field of the $\mathcal{C}$-pulse rotates at a rate given by $\Delta$, and since $\Delta\sim\Omega_0$ the spin cannot adiabatically follow the microwave field~\cite{tewari_berrys_1989}. Also, in both sequences, the spin does not start in an eigenstate of the microwave field operator~\cite{zeng_berry_1995}. For Sequence 1, $\phi_\text{dyn} = 0$ as a result of the Hamiltonians defined in Eq. \ref{eq:h2}, though the dynamic phase reappears when the initial state is swapped so that the NV spin begins in the $\ket{+1}$ state (parallel to the $+z$-axis, rather than antiparallel for $\ket{0}$). 

Isolating the geometric and dynamic phases in the general setting is of interest, since this is the key to exploiting the purported benefits of geometric and topological phases. In the experimental cases we focus on, this is a non-trivial procedure. This can be contrasted to the example of a Berry phase, which could be accumulated by adiabatically rotating ($\Delta\ll\Omega_0$) the microwave field about the $z$-axis, for instance. Here, reversing the cyclic variation of the Hamiltonian, in the form of the rotation direction of the microwave field, inverts the sign of the geometric phase but leaves unchanged the dynamic phase. A spin-echo pulse sequence~\cite{hahn_spin_1950} can then be employed to measure only the geometric phase~\cite{jones_geometric_2000,leek_observation_2007, zhang_experimental_2016, arai_geometric_2018}. In the trajectories considered here, reverse evolution on the cone circuit can be achieved only by a second pulse with inverted detuning, which from Eqs. \ref{eq:faa} also inverts the dynamic phase. 

Many of the peculiarities encountered with the decomposition into geometric and dynamic phases stems from our description of state manipulations in a rotating frame. For instance, in the case of a pure spin-1/2 system initialized into a state orthogonal to the driving field and then subjected to a $\mathcal{C}$-pulse, the dynamic phase vanishes because the spin state evolves on a geodesic of the Bloch sphere, leading to $\phi_\text{AA} = \pi=\phi_T$. Due to the energy level structure of the NV triplet (Hamiltonians Eq.\ref{eq:h2}), such a relation concerning zero dynamic phase no longer hold true: we calculate zero dynamic phase for Sequence 1, instead. The physical significance of zero dynamic phase will become apparent when we consider multiple sequential evolutions applied to the state vector, as only for the case $\phi_\text{dyn} = 0$ can we use simple geometric arguments to predict the system dynamics.

\section{Composite trajectories}\label{sec:comp}

\begin{figure}
	\centering
		\includegraphics[width = \columnwidth]{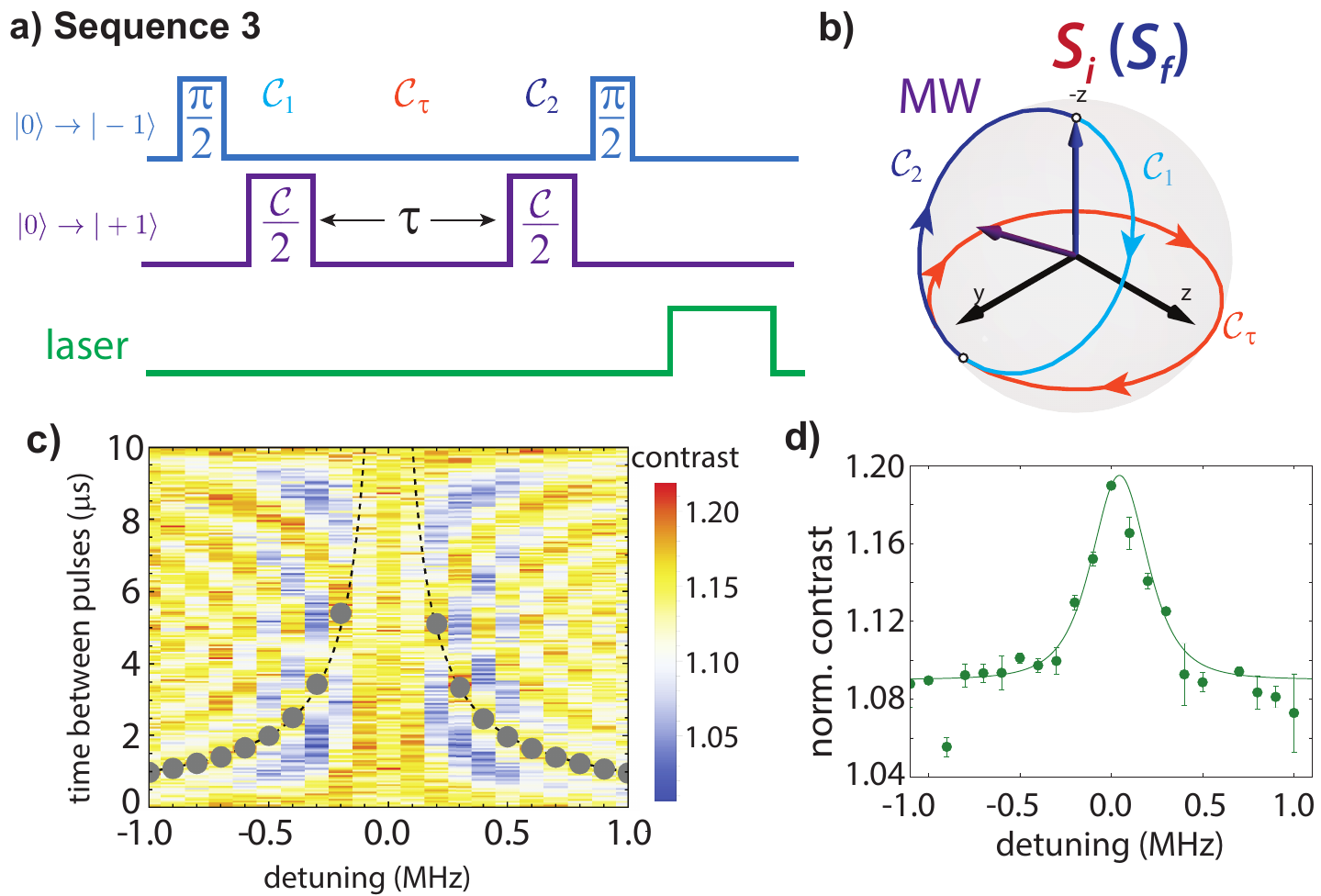}
	\caption{Composite cyclic evolution pulse sequence, Bloch sphere trajectories, free-evolution fringes and results. (a) Sequence 3 consists of two $\mathcal{C}/2$-pulses acting on the $\{0,+1\}$ subsystem with detuning $\Delta$, separated in time by $\tau = 1/\Delta$. A resonant Ramsey sequence applied to the $\{0,-1\}$ subspace reads out the total phase accumulated over the three geometric stages $\mathcal{C}_1,\mathcal{C}_\tau$ and $\mathcal{C}_2$ traced by the spin vector on the Bloch sphere (note inverted $z$-axis) (b). (c) We first verify that the spin vector precesses at a rate $\Delta$ by measuring interference fringes resulting from application of only the $\{0,+1\}$ pulses of Sequence 3 and varying the time between pulses. Gray points depict the fitted period, dashed lines denote $1/\Delta$. (d) Applying Sequence 3 to the NV spin yields data consistent with application of a single $\mathcal{C}$-pulse, indicating the the geometric nature of the free precession stage $\mathcal{C}_\tau$ is suppressed by an accompanying dynamic phase. Error bars derive from standard error from three repeated measurements, solid lines denote fits.}
	\label{fig:fig4}
\end{figure}

We now consider the case of multiple sequential trajectories, under different Hamiltonians. We consider first the case of Sequence 3, depicted in Fig. \ref{fig:fig4}(a). A resonant Ramsey sequence interrogates the $\{0,-1\}$ subspace while a second sequence consisting of two half-duration $\mathcal{C}$-pulses ($\mathcal{C}/2$-pulses) is applied to the $\{0,+1\}$ subspace. The $\mathcal{C}/2$-pulses, with detuning $\Delta$, are separated by a free evolution period of $\tau = 1/\Delta$. Exactly like conventional Ramsey interferometry, the state vector precesses at $\Delta$ in the frame where the microwave field of the $\mathcal{C}/2$-pulses appears stationary.  We experimentally verify that the $\mathcal{C}/2$-pulses initiate free precession by applying them in analogy to a simple Ramsey experiment to a pure $\ket{0}$ state without any reference interferometer, and varying the free evolution time and measuring the period of resulting fringes (Fig.\ref{fig:fig4}c). For $\mathcal{C}/2$-pulses close to resonance (where $\tau\rightarrow\infty$), we cap the free precession time at $10\,\upmu$s. 

As depicted in Fig. \ref{fig:fig4}, there are three geometric manipulations applied to the spin vector, which begins longitudinally aligned to the $-z$-axis. $\mathcal{C}_1$ and $\mathcal{C}_2$ constitute the half-cones subtended by the two $\mathcal{C}/2$-pulses, while $\mathcal{C}_\tau$ denotes the path traversed during free evolution. The geometric component of the total phase accumulated over path $\mathcal{C}_\tau$, identified by the solid angle swept by the state vector in the free-evolution stage (where the spin precesses at $\Delta$ in the resonant frame) is dependent on the detuning of the $\mathcal{C}/2$-pulse applied to the longitudinal spin. The free precession observed in Fig \ref{fig:fig4}(c) confirms that the spin vector precesses in the frame where the microwave field is stationary. However, the results shown in Fig. \ref{fig:fig4}(d) are consistent with that of a single $\mathcal{C}$-pulse being applied to the $\{0,+1\}$ subspace (\emph{e.g.} Fig. \ref{fig:fig2}(c)). The effects of the geometric phase accumulated during the free evolution $\mathcal{C}_\tau$, which depends on the detuning $\Delta$ of the initiating $\mathcal{C}$-pulse, but not on the free precession rate, is not observed.

\begin{figure}
	\centering
		\includegraphics[width = \columnwidth]{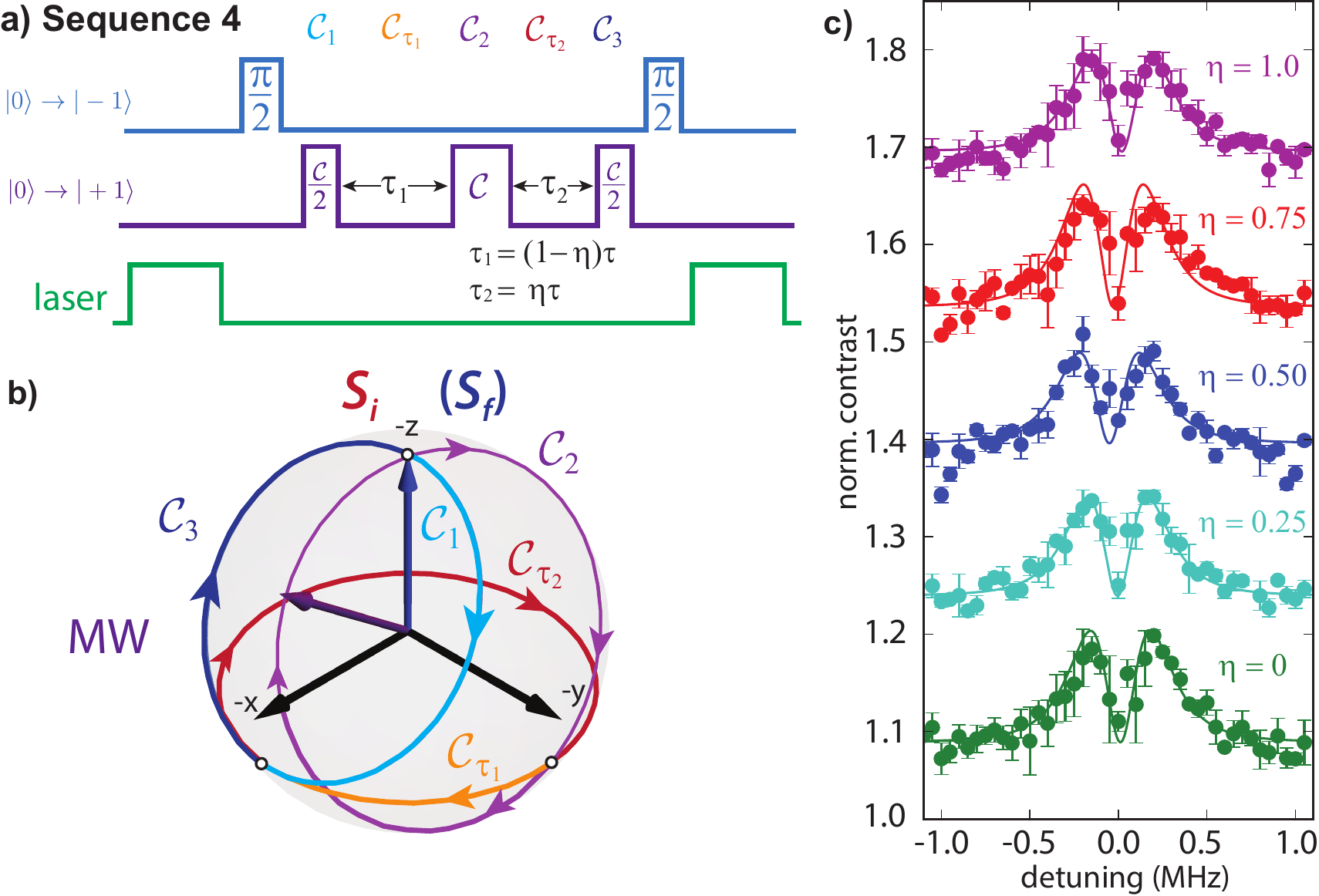}
	\caption{(a) Sequence 4 is identical to Sequence 3, but with a full $\mathcal{C}$-pulse applied after a time $\tau_1 = (1-\eta)\tau$, followed by further free evolution $\tau_2 = \eta\tau$ such that $\tau_1+\tau_2 = 1/\Delta$, with $\eta$ a parameter that determines where in the free evolution the $\mathcal{C}$-pulse is applied. (b) Paths taken by the spin vector under each geometric manipulation step depicted on the Bloch sphere, with the frame rotation rate set by the microwave field driving $\mathcal{C}$-pulses, note inverted $z$-axis. $\mathcal{C}_i$ denotes evolution under the microwave fields and $\mathcal{F}_i$ denotes free-evolution steps. (c) Experimental results for various values of $\eta$, which determines where in the free-evolution step the complete $\mathcal{C}$-pulse is applied. The output of the complete interferometry sequence is identical to two $\mathcal{C}$-pulses applied consecutively.}
	\label{fig:fig5}
\end{figure}

As a further demonstration, we consider Sequence 4, depicted in Fig. \ref{fig:fig5}(a), where a complete $\mathcal{C}$-pulse is inserted at a time $\tau_1$ after the first $\mathcal{C}/2$-pulse, executing a complete rotation of the spin vector over a path denoted $\mathcal{C}_{12}$, the geometry of which is determined by $\tau_1$. A further free evolution time $\tau_2$ follows with $\tau_1+\tau_2 = 1/\Delta$ before the spin returns to the longitudinal spin state. We therefore have a composite path where the geometric phase accumulated in one stage depends on the preceding stage, \emph{i.e.} $\tau_1$ determines the solid angle swept during $\mathcal{C}_{2}$. The results of this sequence, for various ratios $\tau_1/\tau_2$, are plotted in Fig. \ref{fig:fig5}. In a similar vein to Sequence 3, the results are the same regardless of $\tau_1$ and $\tau_2$, and are equivalent to that of two $\mathcal{C}$-pulses applied consecutively. The configuration of the spin vector when the $\mathcal{C}$-pulse is applied, and hence the free evolution stage, appears to have no bearing on the resulting phase accumulation. Such an observation would be expected if no free evolution occurred, though this would infer that the time between the $\mathcal{C}/2$-pulses is immaterial, at odds with the data in Fig. \ref{fig:fig4}(c). 

The resolution of this supposed paradox lies in careful application of Eq. \ref{eq:aasa} to decompose the total phase accumulated over each segment of the trajectory into dynamic and geometric parts for both Sequences 3 and 4. Beginning with Sequence 3, we consider the application of Eq. \ref{eq:aasa} to the free-precession trajectory $\mathcal{C}_\tau$ for the state immediately following the $\mathcal{C}/2$-pulse, 
\begin{eqnarray}
\ket{\psi_{\mathcal{C}/2}} &=& U_{\mathcal{C}/2}\ket{\psi(t=0)}\nonumber\\
                      &=& -e^{i\phi_T/2}\left(\cos\frac{\theta}{2}\ket{+1} +  \sin\frac{\theta}{2}\ket{0} \right),\label{eq:sv}
\end{eqnarray}
with $U_{\mathcal{C}/2}$ a unitary operator that executes a $\mathcal{C}/2$-pulse, $\cos\frac{\theta}{2} = \Omega/\sqrt{\Delta^2+\Omega^2}$, $\sin\frac{\theta}{2} = \Delta/\sqrt{\Delta^2+\Omega^2}$ and $\phi_T = \pi(1+\Delta/\sqrt{\Delta^2+\Omega^2})$. Here, $\theta$ is the angle the Bloch vector makes to the $z$-axis. Computing the dynamic and geometric phases accumulated during the free evolution step, we find
\begin{subequations}
\begin{eqnarray}
\phi_\text{dyn} &=&  \pi(1+\cos\theta)\\
\phi_\text{AA}  &=&  -\pi(1+\cos\theta) \\
                &=& -\phi_\text{dyn},\nonumber
\end{eqnarray}
\end{subequations}
with the resulting total phase $\phi_\text{dyn}+\phi_\text{AA} = 0$. While the geometric and dynamic phases depend on the detuning of the preceding $\mathcal{C}/2$-pulse, the total phase accumulated during the free evolution step is zero, and thus the results of Sequence 3 are consistent with the application of a single $\mathcal{C}$-pulse. Calculating the solid angle $\Theta$ swept out by the spin vector at the centre of the Bloch sphere using Eq. \ref{eq:solidangle} yields $\Theta = 2\pi(1-\cos\theta)$, \emph{i.e.} $2\phi_\text{AA}+2\pi$. The solid angle spanned by the state vector on the Bloch sphere retains its relationship to the Aharonov-Anandan phase. However, an `unseen' accompanying dynamic phase is present that, in this case, exactly cancels the AA phase. Unlike simple, archetypal trajectories commonly invoked to demonstrate geometric phases (like Sequence 1, and numerous other examples~\cite{zwanziger_berrys_1990}), simple geometric arguments are no longer indicative of the actual measured dynamics of the system. 

Analysis of Sequence 4 yields an analogous result, with the initial state vector immediately after $\tau_1$ given by   
\begin{eqnarray}
\ket{\psi_\text{free}} =& U_\text{free}(\tau_1) U_{\mathcal{C}/2}\ket{\psi(t=0)}\nonumber\\
                     =& - e^{i\phi_T/2}\left(e^{i\Delta\tau_1}\cos\frac{\theta}{2}\ket{+1}+ \sin\frac{\theta}{2}\ket{0} \right),\label{eq:sv2}
\end{eqnarray}
with $U_\text{free}(\tau_1)$ the free evolution operator for time $\tau_1$. We now consider the trajectory executed when the $\mathcal{C}$-pulse is applied. The transformation $|\tilde{\psi}(t)\rangle = e^{-if(t)}|\psi(t)\rangle$ yields $|\tilde{\psi}(t_{2\pi})\rangle = |\tilde{\psi}(0)\rangle$ for $f(t) = t/2\left(\Delta+\sqrt{\Delta^2+\Omega^2}\right)$, $f(t_{2\pi})-f(0) = \phi_T$. Evaluating the dynamic and geometric phases, we find

\begin{subequations}
\begin{eqnarray}
\phi_\text{dyn} =& \pi\left(\frac{\Delta}{\sqrt{\Delta^2+\Omega^2}}+\frac{\Delta\cos\theta-\Omega\cos(\Delta\tau_1)\sin\theta}{\sqrt{\Delta^2+\Omega^2}}\right) ,\\
\phi_\text{AA} =& \pi\left(1-\frac{\Delta\cos\theta-\Omega\cos(\Delta\tau_1)\sin\theta}{\sqrt{\Delta^2+\Omega^2}} \right).
\label{eq:faa2}
\end{eqnarray}
\end{subequations}

The resulting total phase is $\phi_T  = \pi\left(1+\Delta/\sqrt{\Delta^2+\Omega^2}\right)$, independent of the preceding free-evolution time, and when combined with the results of Seq. 3 we see that the total phase shift imparted by Seq. 4 is simply $2\phi_T$: two $\mathcal{C}/2$-pulses and one $\mathcal{C}$-pulse, consistent with the data depicted in Fig. \ref{fig:fig5}. We can confirm the relationship between the solid angle swept out by the spin vector and the geometric phase, using Eq. \ref{eq:solidangle}, yielding $\Theta = 2\phi_\text{AA}$. The results discussed in this Section show that despite retaining the geometric interpretation of solid angles enclosed by paths on the Bloch sphere, the presence of an inseparable dynamic phase renders purely geometric inference of the system behavior insufficient. The elimination of the dynamic phase for Sequence 3 and 4 is non-trivial, as time-reversal operations like spin-echo (backward evolution along the geometric paths) will also change the sign of the dynamic phase, which will then add rather than cancel. 

\section{Discussion}\label{sec:disc}
The presence of a dynamic phase that accompanies the geometric phase accumulation precludes a purely geometric consideration of the system dynamics. Additionally, the NV Hamiltonian yields an energy spectrum that makes the geometric phase accumulated in a pseudospin-1/2 subspace different to that observed in a pure spin-1/2 system. We will now discuss these findings in detail.

While the application of Eq. \ref{eq:aap} to Sequence 1 points to a zero dynamic phase for that particular state manipulation, this is purely a result of the expression used for the Hamiltonian, Eq. \ref{eq:h2}. Changing the initial state to $m_S = +1$ rather than $m_S = 0$ would yield a non-zero dynamic phase from Eq. \ref{eq:aap}, despite the two cases being basically identical geometric operations. A natural question to consider is the significance of a purely geometric phase in the context of the Aharonov-Anandan phase. In the adiabatic limit, experiments have provided some evidence that geometric phase is more resilient to certain forms of noise than dynamic phase~\cite{berger_exploring_2013}. A comparably exhaustive analysis of the noise resilience properties or otherwise of the non-adiabatic Aharonov-Anandan phase is beyond the scope of this work, though we note that adiabaticity plays an important role in reported improved noise characteristics of the Berry phase~\cite{de_chiara_berry_2003}.

This work depicts slowly-evolving ($\sim 1/\Omega$) spin vectors on the Bloch sphere, which ignores the fact that the spin is accumulating dynamic (Larmor) phase at a much faster rate in a non-rotating frame. Geometric phases expressed in a transformed frame of reference, such as the microwave rotating frame in this work, are often called `frame-dependent phases'~\cite{kendrick_invariance_1992,skrynnikov_geometric_1994}. The concept of a frame-dependent geometric phase offers some more perspective on the significance of `purely' geometric phases in our work. An analogy with an optical experiment probably best illustrates the point. In Refs. \cite{chiao_observation_1988, jiao_two_1989} the system under consideration is a non-planar Mach-Zehnder interferometer (MZI), designed such that the propagation direction of the light (wavevector) executes a closed circuit in three-dimensional space. In this case, the dynamic phase is identified with the overall path length of each interferometer arm, and the geometric phase with the solid angle swept out by opposite-direction geometric circuits on each arm. Canceling the dynamic phase amounts to ensuring each interferometer arm has the same path length, so the relative phase measured at the output port contains only the geometric phase.

In our case, a similar identification applies, in that each interferometer arm is analogous to one of the two spin-1/2 subspaces of the NV ground state. The natural extension to make subsequently is that the dynamic phase is therefore related to the Larmor phase, \emph{i.e.} precession at $\sim2.870\,\text{GHz}\pm\gamma\,B_0$. However, the entirety of this `Larmor' dynamic phase is accounted for by working in a frame rotating at the microwave frequency tuned to each transition, thus only phase accumulation \emph{relative} to the angular frequency of \emph{each} microwave field is detected in experiments. Carrying the analogy of separate interferometer arms further, the equivalent case for the non-planar MZI are zero-length arms for resonant radiation dressing each transition, and when a detuning is applied in order to execute a $\mathcal{C}$-pulse, for instance, this must therefore be accompanied by a non-zero path length in which the spin manipulation takes place - which in our case is entirely by altering the detuning.  

The notion of a frame-dependent geometric phase can be put on a more formal footing by noting there is a distinction between the energy operator, which yields the system's energy eigenvalues, and the Hamiltonian~\cite{kendrick_invariance_1992, kobe_invariance_1990,kobe_exact_1990}, which are different under a unitary transform. Under a unitary transform $U$, the energy operator is $E = UHU^{-1}$, whereas the Hamiltonian is $H' = E -i \dot{U}U^{-1}$. The transformed energy operator determines the transformed dynamic phase, whereas the second term, $-i \dot{U}U^{-1}$, renders the geometric phase invariant to the unitary transform. These unitary transforms are the rotating-frame transformations we make to simplify the interaction Hamiltonian. In our experiments, the detuning $\Delta$ is used to alter the geometric phases in sequences using a $\mathcal{C}$-pulse, this same detuning term appears in the transformed dynamic phase, making separation non-trivial.

The key finding of this work is that the non-adiabatic Aharonov-Anandan phase is accompanied by an inseparable dynamic phase in a microwave-frequency rotating frame that precludes direct measurement of the geometric component of the total phase, even when the geometric paths are widely varied. In the case of pulse sequences that are tailored so that the dynamic phase is always zero, the dynamics of the full system can in principle be predicted based on geometric arguments on the Bloch sphere. However, in our cases these `purely geometric' trajectories are essentially trivial. Examining Sequence 4, and Eqs. \ref{eq:faa2}, we can see that the dynamic phase vanishes for $\tau_1 = 2\pi/\Delta$, which corresponds to the case where the spin executes a $2\pi$ rotation about the $z$-axis and returns to where it was initialised by the $\mathcal{C}/2$ pulse. As we saw from Sequence 3, this free evolution step imparts no measurable effect on the spin vector, since the geometric and dynamic phases always cancel. We thus find ourselves at Sequence 1, being the exemplar case when the dynamic phase vanishes. 

An intriguing direction of further investigation is exploiting the quantum Zeno effect~\cite{facchi_berry_1999}, where repeated projective measurements on one state of the two-level system during a cyclic state manipulation yields a phase that cancels the $\phi_\text{dyn}$ accumulated during a $\mathcal{C}$-pulse. Recently demonstrated for a two-level spin system formed from a $^{87}$Rb Bose-Einstein condensate~\cite{do_measuring_2019}, the quantum Zeno scheme yields a measured phase consistent with the purely geometric AA phase. Our work presents an ideal opportunity to further evaluate the Zeno scheme and reveal purely geometric quantum state evolution. We note with some interest the very similar functional forms of geometric and dynamic phase we have studied in this work, and leave open the question as to whether the Zeno scheme can isolate a purely geometric phase in examples such as Sequences 3 and 4.        

\section{Conclusions}
In this work, we have studied spin manipulation sequences applied to a pseudospin-1/2 subspace of the spin-1 NV ground state. The simple case of a spin-echo sequence served as an interesting initial starting point, and with the results of Sequence 4 in mind, we immediately see parallels between the two. Changing the duration of the spin-echo sequence necessarily varies the geometric path executed by the spin on the Bloch sphere, whilst the overall trajectory remains cyclic. The overall measured result will not change, and interpretation of the spin echo sequence using the theory of Aharonov and Anandan is unnecessary in the case of an isolated pseudospin-1/2 subspace. Double-quantum pulse sequences ~\cite{mamin_multipulse_2014, bauch_ultralong_2018} that use the full spin-1 ground state of the NV are increasingly being used in the effort to render NV magnetometry insensitive to the deleterious effects of temperature and crystal strain sensitivity. As we have seen in this work, a more nuanced approach to familiar spin manipulations in these circumstances is necessary to understand the system dynamics in detail. 

Geometric phase in both the adiabatic and non-adiabatic regime has been recently studied as an alternative to traditional Ramsey magnetometry~\cite{arai_geometric_2018}, which exhibits a $2\pi$ ambiguity in measured phase. It would be interesting to examine this scheme in the context of our work on the AA phase, and determine if adiabaticity or geometry is more important for noise resilience or decoherence.  Additional avenues of future investigation could focus on the effects of physical rotation on spin manipulations, recently reported in Ref. \cite{wood_observation_2020}. In that particular case, physical rotation of the NV qubit leads to non-linear accumulation of a relative phase between the NV and microwave field, which is detected in a spin-echo experiment. The scheme described here could be implemented to detect the Aharonov-Anandan phase accumulating from rotation of the NV, akin to proposals to detect Berry phase from physical rotation~\cite{maclaurin_measurable_2012}. 

\section*{Acknowledgements}
We thank L. P. McGuinness for provision of the diamond sample. L. C. L. H. acknowledges support from the ARC Centre of Excellence scheme (CE170100012). M.B.P. is supported by the ERC Synergy Grant HyperQ, the EU via project AsteriQs, the BMBF via NanoSpin and DiaPol and a DFG Kosseleck award. This work was supported by the Australian Research Council Discovery Scheme (DP150101704, DP190100949).

\end{document}